\documentclass{IEEEtran}
\usepackage[boxruled]{algorithm2e}
\usepackage{enumerate}
\usepackage{cite}
\usepackage{graphicx}
\usepackage{psfrag}
\usepackage{subfigure}
\usepackage{url}
\usepackage{amsmath}
\usepackage{amsthm}
\usepackage{flushend}
\usepackage{array}
\usepackage{algorithm2e}
\usepackage{amssymb}
\usepackage{amsfonts}
\usepackage{float}
\newcommand\undermat[2]{%
  \makebox[0pt][l]{$\smash{\underbrace{\phantom{%
    \begin{matrix}#2\end{matrix}}}_{\text{$#1$}}}$}#2}
\makeatletter
\renewcommand{\boxed}[1]{\text{\fboxsep=.2em\fbox{\m@th$\displaystyle#1$}}}
\makeatother

\newtheorem{proposition}{Proposition}
\newtheorem*{conjecture*}{Conjecture}
\newtheorem{lemma}{Lemma}

\newtheorem{corollary}{Corollary}
\newtheorem{remark}{Remark}
\newtheorem{definition}{Definition}
\newtheorem{theorem}{Theorem}
\newtheorem{example}{Example}

\title{Index Coding: Rank-Invariant Extensions}
\begin{document}
\author{
\IEEEauthorblockN{Vamsi Krishna Gummadi, Ashok Choudhary, Prasad Krishnan\\}
\IEEEauthorblockA{Signal Processing and Communications Research Centre,\\
International Institute of Information Technology, Hyderabad.\\
Email: \{vamsi.gummadi,ashok.choudhary,prasad.krishnan\}@iiit.ac.in\\}
}
\date{\today}
\maketitle
\thispagestyle{empty}	
\pagestyle{empty}

\begin{abstract}
An index coding (IC)  problem consisting of a server and multiple receivers with different side-information and demand sets can be equivalently represented using a fitting matrix. A scalar linear index code to a given IC problem is a matrix representing the transmitted linear combinations of the message symbols. The length of an index code is then the number of transmissions (or equivalently, the number of rows in the index code). An IC problem ${\cal I}_{ext}$ is called an extension of another IC problem ${\cal I}$ if the fitting matrix of ${\cal I}$ is a submatrix of the fitting matrix of ${\cal I}_{ext}$. We first present a straightforward $m$\textit{-order} extension ${\cal I}_{ext}$  of an IC problem ${\cal I}$ for which an index code is obtained by concatenating $m$ copies of an index code of ${\cal I}$. The length of the codes is the same for both ${\cal I}$ and ${\cal I}_{ext}$, and if the index code for ${\cal I}$ has  optimal length then so does the extended code for ${\cal I}_{ext}$. More generally, an extended IC problem of ${\cal I}$ having the same optimal length as ${\cal I}$ is said to be a \textit{rank-invariant} extension of ${\cal I}$.  We then focus on $2$-order rank-invariant extensions of ${\cal I}$, and present constructions of such extensions based on involutory permutation matrices.   
\end{abstract}
\section{Introduction}

The typical setting in index coding (IC), introduced in \cite{BiK}, is a noiseless broadcast channel where a single source is communicating with a number of receivers, each of which have a set of demands and possess a subset of known source symbols called side-information. The goal of index coding is then to design a transmission scheme which minimises the transmission rate (called as `length' of the index code for the scalar linear index coding case) while satisfying the demands of all the receivers. Different classes of the IC problem were studied based on the configuration of the side-information symbols and the demands. In \textit{unicast IC} \cite{BBJK}, the demand sets of the receivers are disjoint. 
A unicast IC problem can be represented using a graph known as the side-information graph. A number of researchers (see for example \cite{BBJK,DSC}) have obtained bounds on the optimal length of unicast index codes using several graph quantities like chromatic number, independence number, etc. Similar approaches to the uniprior class of IC problems (where demand sets are disjoint) was taken in \cite{OnH,MaK}, and general IC problems were discussed in \cite{TDN}. 
Many of these papers also lead to constructions of (scalar and vector) linear index codes. Linear codes however are not always found to be optimal \cite{LuS}.

An interesting line of work in index coding discussed in prior work such as \cite{BVR,VaR,EbS,ArK1,ArK2} is the characterisation of optimal rates of new `bigger' IC problems based on other smaller `component' problems for which optimal code-lengths are known. For example, in \cite{BVR}, a lifting construction is provided by which a single-unicast problem ${\cal I}_{ext}$ containing $mK$ messages is created from a single unicast problem ${\cal I}$ containing $K$ messages, and it is shown that optimal scalar linear codes for ${\cal I}$ can be lifted to optimal codes for ${\cal I}_{ext}$ having the same length. Using this result, the authors of \cite{BVR} and \cite{VaR} construct optimal scalar and vector linear index codes for several classes of IC problems starting from simpler, more basic, IC problems. In \cite{EbS}, the existence of a graph homomorphisms between complements of side-information graphs of two unicast problems were used to show that the optimal lengths of the two are related. In \cite{ArK1,ArK2}, graph products were used to relate the optimal index coding rates of unicast IC problems with larger side-information graphs with the optimal rates of their component problems with smaller side-information graphs. 

In this work, we look at characterising extensions of a given general IC problem, which have the same optimal length as the given problem. The study of such extensions could be useful in a variety of scenarios. For example, in a dynamic broadcast network, we could have receivers and messages which are being added to the network. In such scenarios, it would be pertinent to enquire into what configurations of side-information should be present in the users so as to maintain the rate of transmission. Index coding is also directly related to topological interference management in wireless networks \cite{Jaf} where such dynamic scenarios can be expected, so this is a problem of practical interest too. Besides, index coding is also related in a dual sense to the distributed storage problem \cite{MAZ}. Optimal index codes thus translate into optimal locally recoverable distributed storage codes (with the largest dimension). Thus, rank-invariant extensions of IC problems could provide insights into doubling constructions of optimal distributed storage codes  (i.e., roughly doubling the dimension as well as the length of the code), while maintaining repair capabilities. The contributions and organization of this work are as follows.
\begin{itemize}
\item After briefly reviewing the basics of index coding, we introduce the notion of an $m$-order extension of an IC problem and also define rank-invariant extensions as those extensions which have the same optimal length as the basic IC problem. We present a straightforward $m$-order extension which is also rank-invariant. (Section \ref{sec1})
\item Focusing on a special type of index codes for the extended IC problems, we obtain some sufficient conditions for rank-invariability on the structure of the basic IC problems and that of their $2$-order extensions. (Section \ref{sec3})
\item Using the sufficient conditions developed in Section \ref{sec3}, we obtain classes of IC problems which are $2$-order rank-invariant extendable and demonstrate their extensions explicitly. (Section \ref{sec4}). 
\end{itemize}
\textit{Notations:} For some positive integer $m$, We denote the set $\{1,2,...,m\}$ as $[1:m]$. The rank of a matrix $A$ is denoted by $rank(A)$. The identity matrix is denoted by $I$ (the size should be clear from the context).

\section{Index Coding via Fitting Matrices} 
\label{sec1}
Formally, the index coding (IC) problem ${\cal I}$ (over some finite field $\mathbb{F}_q$) consists of a source, a set of $m$ receivers, a broadcast channel which can carry symbols from ${\mathbb F}_q$, along with the following.
\begin{itemize}
\item Source message symbols ${\cal X}=\{x_i, i\in[1:K]\}$, each of which is modelled as a $d$-length vector over ${\mathbb F}_q$. We refer to the indices $i\in[1:K]$ as the \textit{messages}, and the symbols $x_i$s as the \textit{message symbols}.
\item For each receiver $t$, a set $D(t)\subseteq {\cal X}$ denoting the set of messages demanded by the receiver $t$. We denote the total number of demands $\sum_{j=1}^m|D(t)|$ by $L$ (we assume that $L\geq K$ as each message is demanded at least once).
\item For each receiver $t$, a set $S(t)\subseteq {\cal X}\backslash D(t)$ denoting the side-information messages available at the $t^{th}$ receiver. 
\end{itemize}
For a message vector $\boldsymbol{x}\in{\mathbb F}^{Kd}$, the source transmits a $r$-length codeword ${\mathbb E}(\boldsymbol{x})$ (where $\mathbb E:{\mathbb F}^{Kd}\rightarrow {\mathbb F}^r$), such that all the receivers can recover their demands. The map ${\mathbb E}$ is then called a  index code, or simply, an index code for ${\cal I}$. The \textit{transmission rate} of the code is defined as $\frac{r}{d}$. If $d=1$, then the index code is known as a \textit{scalar index code}, else it is known as a \textit{vector index code}. A linear encoding function ${\mathbb E}$ is also called a linear index code. The goal of index coding is to find optimal index codes, i.e., those with the minimum possible transmission rate. For scalar linear index codes, we refer to the quantity $r$ as the \textit{length of the code}, and thus rate optimality translates to minimal length codes. In this paper, we focus on scalar linear index codes. A scalar linear index code for ${\cal I}$ is captured equivalently by a matrix $G_{r\times n}$ (with the transmitted codeword being $G\boldsymbol{x}$). 

Corresponding to an IC problem ${\cal I}$ as in the above definition, we can associate a $L\times K$ matrix $F_X$ known as the fitting matrix \cite{BBJK} corresponding to ${\cal I}$, where each row represents a particular demand at a particular receiver, and the columns represent the messages. The matrix $F_X$ is populated as follows. Consider the $l^{th}$ row of the $F_X$ corresponding to demanded message $k$ at a receiver $t$. Then this $l^{th}$ row contains a $1$ at position $k$, a place-holder $X$ at positions indexed by $S(t)$, and $0$s everywhere else. It was shown in \cite{BBJK} that the optimal length of scalar linear codes for ${\cal I}$ is equal to a property of the fitting matrix known as its \textit{minrank}, denoted by $minrk(F_X)$ (the result is shown for unicast problems, but extendable to general problems in a straightforward way). The minrank $minrk(F_X)$ (over ${\mathbb F}_q$) is the minimum rank of the matrix obtained by replacing the $X$s in $F_X$ with arbitrary values from ${\mathbb F}_q$. 

The following lemma is used multiple times in the paper and can be inferred from \cite{BBJK}. The proof follows from the simple observation that each receiver is able to decode its desired symbol by taking linear combinations of the transmitted symbols and the side-information. We leave the details of the proof to the reader.
\begin{lemma}
\label{simplelemma}
Consider an IC problem ${\cal I}$ with $L\times K$ fitting matrix $F_X$. An $r\times K$ matrix $G$ is an index code for ${\cal I}$ if and only if there is a ${L \times r}$ matrix $D$ such that $DG \approx F_X$.
\end{lemma}

As the fitting matrix of a given IC problem completely characterises it, we can use matrices with similar properties to directly define IC problems. We therefore redefine the fitting matrix in the following way.
\begin{definition}[Fitting Matrix]
\label{fitting}
For positive integers $L$ and $K$, a $L\times K$ ($L\geq K$) fitting matrix $F_X$ is a matrix consisting of $1$s, $0$s and $X$s such that
\begin{itemize}
\item Each row contains precisely one $1$ and some number of $X$s and $0$s. 
\item For any column $k\in K$, there exists at least one row in $F_X$ with a $1$ at the $k^{th}$ position. 
\end{itemize}
\end{definition}
Note that every fitting matrix $F_X$ as in Definition \ref{fitting} naturally defines an IC problem constructed in the following way.
\begin{itemize}
\item The IC problem consists of source with $K$ messages and $L$ receivers. 
\item Consider a row of $F_X$ in which $1$ appears in the $k^{th}$ position. Corresponding to this row, associate a unique receiver which demands the $k^{th}$ message, having side information precisely as those messages corresponding to the column indices of $X$s in the row.
\end{itemize}

To facilitate our results, we also give the following definitions.%
\begin{definition}
An $X$-fitting matrix $B_X^X$ is a matrix which consists only of $X$s and $0$s. The $X$-fitting matrix obtained from a fitting matrix $F_X$ by replacing the $1$s in $F_X$ by $X$s is denoted by $F_X^X$.
\end{definition}
\begin{definition}
For a fitting matrix $F_X$ (equivalently, for an $X$-fitting matrix $B_X^X$) we say that a matrix $F \approx F_X$ (equivalently, $F\approx B_X^X$) if the set of zero positions of $F_X$ (equivalently, of $B_X^X$) is a subset of zero positions of $F$.
\end{definition}
\subsection{Rank-Invariant Extensions}
We now formally define extensions of a given IC problem, and also rank-invariant extensions. 
\begin{definition}[Extension of an IC problem]
Let ${\cal I}$ be an IC problem  with fitting matrix $F_X$. An IC problem ${\cal I}_{ext}$ is called an extended IC problem of ${\cal I}$ (or simply, an extension of ${\cal I}$) if its fitting matrix $F_{ext,X}$ contains $F_X$ as a submatrix. The extension ${\cal I}_{ext}$ is called an $m$\textit{-order extension} of ${\cal I}$ if $F$ is a $L\times K$ matrix while $F_{ext,X}$ is an $mL\times mK$ matrix.
\end{definition}
\begin{definition}[Rank-Invariant Extensions]
An extension ${\cal I}_{ext}$ with fitting matrix $F_{ext,X}$ of an IC problem ${\cal I}$ with fitting matrix $F_X$ is called rank-invariant if their optimal code lengths are equal, i.e., $minrk(F_{ext,X})=minrk(F_X).$ 
\end{definition}
The following lemma gives a simple lower bound on $minrk(F_{ext,X})$.
\begin{lemma}
\label{lemma_lengthbound}
\[
minrk(F_{ext,X})\geq minrk(F_X)
\]
\end{lemma}
\begin{IEEEproof}
Suppose $F_{ext}$ is a matrix such that $F_{ext}\approx F_{ext,X}$. Let $F$ denote the submatrix of $F_{ext}$ containing the first $L$ rows and first $K$ columns. Thus we must have $rank(F)\leq rank(F_{ext})$. Furthermore note that $F\approx F_X$ by the structure of $F_{ext}$. Since this holds for arbitrary such $F_{ext}$, we have $minrk(F_X)\leq minrk(F_{ext,X})$. 
\end{IEEEproof}
The following theorem describes a straightfoward rank-invariant extension of a given IC problem.
\begin{theorem} 
\label{mainthm1}
Let ${\cal I}$ be an IC problem with $L\times K$ fitting matrix $F_X$ and a index code $G_{r\times K}$. Consider a fitting matrix $F_{{ext,X}}$ for an $m$-order extended IC problem ${\cal I}_{ext}$ of ${\cal I}$ obtained from $F_X$ as follows.
\begin{itemize}
    \item The row-indices of $F_{ext,X}$ is divided into $m$ blocks (each block is of size $L$) and the column indices into $m$ column-blocks (each of size $K$). For $i,j\in [1:m]$, the $(i,j)^{th}$ block in $F_{ext,X}$ is $F_X$ if $i=j$, and $F_X^X$ otherwise.
\end{itemize}
In other words,
\[
F_{ext,X} = \left(
  \begin{array}{rcrcr}
    F_X & F_X^X & \hdots & \hdots & F_X^X  \\
    F_X^X & F_X&F_X^X&\hdots &\vdots\\
    \vdots&F_X^X&\ddots&\hdots&\vdots\\
    \vdots&\vdots&\vdots&F_X&F_X^X \\
    F_X^X &\hdots&\hdots&F_X^X &F_X \\
  \end{array}
\right).
\]
Then ${\cal I}_{ext}$ has the following index code 
\[
{G}_{ext} = (\underbrace{G|G|...|G}_{m~\text{times}}).
\]
Furthermore, $G_{ext}$ is an optimal code for ${\cal I}_{ext}$ if $G$ is optimal for ${\cal I}$. Thus ${\cal I}_{ext}$ is a rank-invariant extension of ${\cal I}$.
\end{theorem}
\begin{IEEEproof}
By Lemma \ref{simplelemma}, we only have to demonstrate some $mL\times r$ matrix $D_{ext}$ such that $D_{ext}G_{ext}\approx F_{ext,X}$. Let $D_{ext}=(\underbrace{D|D|....|D}_{m~\text{times}})^T$. Then,
 \[
 D_{ext}G_{ext}= \begin{pmatrix}
DG&DG&\hdots&DG\\
DG&DG&\hdots&DG\\
\vdots&\vdots&\hdots&\vdots\\
DG&DG&\hdots&DG
\end{pmatrix}\approx F_{ext,X},
 \]
where the last relation holds because $DG\approx F_X$ (and thus $DG\approx F_X^X$ as well). Finally we prove optimality. By Lemma \ref{lemma_lengthbound}, the optimal length of a code for ${\cal I}_{ext}$ is greater than or equal to the optimal length of a code for ${\cal I}$. However the lengths of the codes $G_{ext}$ and $G$ are the same. Hence $G_{ext}$ is optimal for ${\cal I}_{ext}$. This proves the theorem. 
\end{IEEEproof}
\begin{remark}
In Theorem \ref{mainthm1}, if ${\cal I}$ is a unicast problem, then so is ${\cal I}_{ext}$. This case was handled in \cite{BVR} (see Theorem 1 of \cite{BVR}). It turns out that the unicast result from \cite{BVR} can also be seen as a special case of Theorem 5 from \cite{ArK2} and Corollary 1 from \cite{EbS} (we leave the details to the reader). The general IC problem presented in Theorem \ref{mainthm1} here is however is not handled in any of these papers. Furthermore we also specifically focus on rank-invariant extensions, which has not been considered before in literature. 
\end{remark}
\section{Sufficient conditions for rank-invariance for a class of $2$-order extensions}
\label{sec3}
In Theorem \ref{mainthm1}, we obtained a particular method of obtaining a $m$-order extension which is rank-invariant. In the rest of this paper, we focus on $2$-order rank-invariant extensions. In general, it seems to be hard to directly characterise all possible $2$-order rank-invariant extensions of an arbitrary given problem. Therefore we approach the problem in the reverse direction in the following sense 
\begin{itemize}
\item $\mathbf{Q1}:$ What can be the properties of ${\cal I}$ and ${\cal I}_{ext}$, given that an index code for ${\cal I}$ can be extended to a same-length index code for ${\cal I}_{ext}$?
\end{itemize}
We thus focus on particular types of problems which are `extendable' to particular kinds of $2$-order rank-invariant extensions. In this section, we look at some sufficient conditions towards obtaining such extendable IC problems (and their extensions) based on the properties of their index codes. 

The following lemma assumes a particular structure on the index code of ${\cal I}_{ext}$ towards answering $\mathbf{Q1}$. 
\begin{lemma}
\label{identity}
Suppose 
\[
G'=\left(
   \begin{array}{c|c}
    G & A \\
    \hline
    A & G 
    \end{array}
\right)
\]
is an index code to an index coding problem, with $A$ and $G$ both $r\times K$ matrices such that $rank(G')=rank(G)=r$. Then $A$ = $CG$ where $C$ is some matrix such that $C^2$ = $I$ (i.e., $C=C^{-1}$). 
\end{lemma}
\begin{IEEEproof}

Note that $(G~A)$ is a rank $r$ matrix. Thus we must have for $i,j\in[1:r]$,
\[
\sum_{i=1}^r\alpha_{i,j}(g_i,a_i) = (a_j,g_j),
\]
for some scalars $\alpha_{i,j} \in {\mathbb F}_q$, where $(g_i,a_i)$ represents the $i^{th}$ row of $(G~A)$ and $(a_j,g_j)$ is the $j^{th}$ row of $(A~G)$. We thus have
\begin{align}
\label{eqn5}
\sum_{i=1}^r\alpha_{i,j}g_i = a_j,   ~~\sum_{i=1}^r\alpha_{i,j}a_i = g_j.
\end{align}
Let $C$ = $[\alpha_{i,j}], \forall i,j\in[1:r]$ (i.e., $C$ is the $r\times r$ matrix with $\alpha_{i,j}$ being the $(i,j)^{th}$ element). By (\ref{eqn5}) we have $CG$ = $A$ and $CA$ = $G$. Therefore $C^2$ = $I$.
\end{IEEEproof}
\begin{remark}
Lemma \ref{identity} means that $(A~G)$ and $(G~A)$ are both equivalent codes for the given IC problem. Thus it can be looked at as dealing with problems which are a simple generalization of $2$-order rank-invariant extension in Theorem \ref{mainthm1}. Using $A=G$, we get the extension ${\cal I}_{ext}$ given in Theorem \ref{mainthm1} as one possible extension having a code $(G~G)$).
\end{remark}
Self-inverse matrices, for instance the matrix $C$ from Lemma \ref{identity}, are also called as \textit{involutory matrices} in literature (for example, see \cite{BrG}). By Lemma \ref{identity}, suppose $\begin{pmatrix}
G & CG \\
CG & G
\end{pmatrix}$ is an index code to ${\cal I}_{ext}$ while $G$ is a solution of ${\cal I}$. What kind of relationship can we expect between ${\cal I}$ and ${\cal I}_{ext}$? Theorem \ref{thmdcgbx} that follows is a first step towards answering this question, and will be successively refined in our later results in the next section.
\begin{theorem}
\label{thmdcgbx}
Consider the IC problem ${\cal I}$ with $L\times K$ fitting matrix $F_X$ and index code $G_{r\times K}$. Let $D_{L\times r}$ be a matrix such that $DG \approx F_X$ (such a $D$ exists by Lemma \ref{simplelemma}). For some $r\times r$ involutory matrix $C$, let $B_X^X$ be an $L\times K$ $X$-fitting matrix such that $DCG \approx B^X_X$. Then the $2$-order extended IC problem ${\cal I}_{ext}$ with fitting matrix
$ \begin{pmatrix} 
F_X & B_X^X \\
B_X^X & F_X 
\end{pmatrix}$ has the  index code given by $(G~~CG)$. Furthermore, if $G$ is optimal for ${\cal I}$, then so is $(G~~CG)$ and hence ${\cal I}_{ext}$ is a rank-invariant extension of ${\cal I}$.
\end{theorem}

\begin{IEEEproof}
Note that, by the given conditions,
\begin{equation*}
\begin{aligned}
\begin{pmatrix}
D & \boldsymbol{0} \\
\boldsymbol{0} & D
\end{pmatrix}
\begin{pmatrix}
G & CG \\
CG & G
\end{pmatrix}
\approx
\begin{pmatrix} 
F_X & B_X^X \\
B_X^X & F_X 
\end{pmatrix},
\end{aligned}
\end{equation*}
where $\boldsymbol{0}$ is a zero-matrix of appropriate size. By Lemma \ref{identity},
$\begin{pmatrix}
G & CG \\
CG & G
\end{pmatrix}$
is a solution to 
$\begin{pmatrix} 
F_X & B_X^X \\
B_X^X & F_X 
\end{pmatrix}$.
As  $\begin{pmatrix}
G & CG \\
CG & G
\end{pmatrix}$ is a rank $r$ matrix, we thus have that $(G~~CG)$ is a  index code to the IC problem.  The optimality follows by similar arguments as Theorem \ref{mainthm1}.
\end{IEEEproof}
\begin{remark} Note that if we fix $C=I$, and $B_X^X=F_X^X$ (the $X$-fitting matrix corresponding to $F_X$) in Theorem \ref{thmdcgbx}, we recover the result in Theorem \ref{mainthm1} for $2$-order extensions.
\end{remark}


%


\section{New Classes of Rank-Invariant $2$-order Extendable IC problems and their extensions}
\label{sec4}
In this section, we present new classes of IC problems which lend themselves to $2$-order extensions which are rank-invariant. Our results will be founded upon Theorem \ref{thmdcgbx}. Towards that end, we first discuss specific involutory matrices which are conducive to our goal. 
\subsection{Involutory Permutation Matrices}
The class of permutation matrices (containing exactly one `1' in each row and column) has a non trivial intersection with the class of involutory matrices (of the same dimension). In the rest of the paper, we focus on involutory permutation matrices, as they enable us to define classes of new rank-invariant index coding problems drawing from Theorem \ref{thmdcgbx}. As the inverse of a permutation matrix is its transpose, we thus have that $C^{-1}=C^T=C$ for a permutation involutory matrix $C$.

A permutation matrix (viewed as a row permutation) can be equivalently represented as a disjoint product of its \textit{cycles} (called the cycle representation, see \cite{Stan} for instance). For example, the permutation matrix
\[
P=\begin{bmatrix}
0 & 1 & 0 & 0 & 0 & 0 \\
0 & 0 & 1 & 0 & 0 & 0 \\
1 & 0 & 0 & 0 & 0 & 0 \\
0 & 0 & 0 & 0 & 0 & 1 \\
0 & 0 & 0 & 0 & 1 & 0 \\
0 & 0 & 0 & 1 & 0 & 0 \\
\end{bmatrix}
\]
has the cycle notation $(132)(46)(5)$, which we denote as $\sigma_P$. 
For a $r\times r$ permutation matrix $P$, we also abuse the notation $\sigma_P$ denote the permutation function on $[1:r]$ (for example, for the above matrix, we have $\sigma_P(1)=3$ indicating that the first row is mapped to the third by pre-multiplication by $P$). It is easy to see that a  permutation matrix is involutory if and only if its cycle representation contains no cycle of length greater than two, and the list of such permutations is known (see \cite{TMu}, for instance). 
For example, the permutation matrix $C$ corresponding to the permutation $\sigma_C=(1 4)(2 3)(5)(6)(7)$ is an involutory matrix as well. An element $i$ in a cycle $(i)$ of size $1$ of a permutation $\sigma_C$ is called a \textit{fixed point} of $\sigma_C$, while we refer to a cycle of size $2$, $(k,\sigma_C(k))$, as a \textit{swap} of $\sigma_C$. We also note that $\sigma_C$ can represent the row permutation corresponding to $C$, as well as the column permutation, as $C=C^T$.

Towards obtaining our results in Section \ref{specialclass}, we require the following observations. Let ${\mathbb M}_r({\mathbb F}_q)$ denote the the matrix ring consisting of all $r\times r$ matrices over ${\mathbb F}_q$ with usual addition and multiplication. The centralizer of an involutory matrix $C$ in the ring ${\mathbb M}_r({\mathbb F})$, denoted by ${\cal C}_{\mathbb M}(C)$, is the set of all matrices in ${\mathbb M}_r({\mathbb F})$ which commute with $C$. Clearly all matrices which are polynomials in $C$ exist in ${\cal C}_{\mathbb M}(C)$. However, since $C^2=I$, the polynomials in $C$ are unique only upto degree $1$ (over ${\mathbb F}_2$ we thus have only two full-rank choices, $I$ and $C$). The following lemma shows another matrix (used in Section \ref{specialclass}) which commutes with $C$. 

\begin{lemma}
\label{anothercommutingmatrix}
Let $C$ be an involutory permutation matrix . Consider a matrix $Y=I+C-C_1$, where $C_1$ is a matrix constructed as follows.
\begin{itemize}
\item For each fixed point $i$ of $\sigma_C$, the $i^{th}$ column of $C_1$ is the same as the $i^{th}$ column of $I$.
\item Every other column of $C_1$ is zero.
\end{itemize} 
Then $Y$ commutes with $C$. 
\end{lemma}
\begin{IEEEproof}
We first note that $I+C$ commutes with $C$. We only have to show that $C_1$ also commutes with $C$. 

Note that $C_1=C_1^T$. Consider the matrix $CC_1$. For any column index $i$ which is a fixed point of $\sigma_C$, the $i^{th}$ column of $CC_1$ is the $i^{th}$ column of $C$ (which itself is the  $i^{th}$ column of $I$). All other columns of $CC_1$ are zero. Thus $CC_1=C_1$. Furthermore this means that 
\begin{align}
\label{eqn30}
C_1C&=C_1^TC^T\\
\nonumber
&=(CC_1)^T\\
\nonumber
&=C_1^T=C_1,
\end{align}
where (\ref{eqn30}) holds because $C=C^T$ and $C_1=C_1^T$. This proves the lemma.
\end{IEEEproof}
%
\subsection{A Simple Rank-Invariant $2$-order extension of any IC problem}
\label{subsecsimpleclass}

Theorem \ref{thm1} below is the main result in this section. It leverages the idea of using an involutory permutation matrix $C$ and the existence of matrices from ${\cal C}_{\mathbb M}(C)$. Let 	
\begin{equation}
\label{eqn411}
G_{r\times K}=[A_1|A_2|...|A_T|P]
\end{equation}
be a rank $r$ index code for an IC problem ${\cal I}$ with $L\times K$ fitting matrix $F_X$ such that each $A_i$ is an $r\times r$ matrix. Let the set of column indices of $A_i$ in $G$ be denoted as ${\cal I}_i$ (it is not necessary that $A_i$s appear in consecutive blocks as shown in (\ref{eqn411}), but the sets ${\cal I}_i$s must be disjoint). We then have the following main theorem of this section.
\begin{theorem}
\label{thm1}
Let $C$ be a $r\times r$ involutory permutation matrix such that $A_i\in {\cal C}_{\mathbb M}(C),$ 
$~\forall i$. Let $B_X^X$ be an $X$-fitting matrix obtained from $F_X$ by the following procedure.
\begin{itemize}
\item For $j=1,...,T$, the subset of columns of $F_X$ indexed by ${\cal I}_j$ is permuted by $\sigma_C$.
\item All elements in the $K-Tr$ columns of $F_X$ not in any ${\cal I}_j$ are replaced by $X$.
\item All the $1$s in the matrix obtained after above two operations are replaced by $X$s. 
\end{itemize} 
Then  $(G~CG)$ is an index code for the extended IC problem ${\cal I}_{ext}$ with fitting matrix 
$\begin{pmatrix} 
F_X & B_X^X \\
B_X^X & F_X 
\end{pmatrix}.
$ Furthermore, if $G$ is optimal for ${\cal I}$ , then $(G~CG)$ is an optimal code for ${\cal I}_{ext}$ and ${\cal I}_{ext}$ is a rank-invariant $2$-order extension of ${\cal I}$.
\end{theorem}
\begin{IEEEproof}
Since $G$ is a solution, there exists a $L\times r$ matrix $D$ such that $[DA_1|DA_2|\hdots|DA_{T}|DP]=DG\approx F_X$.

Now consider the matrix $DCG$. We have,
\begin{align}
\nonumber 
DCG=&DC[A_1|A_2|...|A_T|P],\\
\nonumber
=&D[CA_1|CA_2|...|CA_T|CP],\\
\label{eqn20}
=&[DA_1C|DA_2C|...|DA_TC|DCP],\\
\label{eqn21}
\approx& B_X^X,
\end{align}
where (\ref{eqn20}) holds because $C$ and $A_i$s commute, and (\ref{eqn21}) follows by structure of $B_X$ as stated and because post-multiplication by $C$ permutes the set of column indices by $\sigma_C$. By Theorem \ref{thmdcgbx}, we have our result.
\end{IEEEproof}




Using Theorem \ref{thm1}, we can obtain a simple rank-invariant $2$-order extension of any IC problem. Let ${\cal I}$ be an IC problem with $L$ demands, $K$ messages and with an optimal index code $G'_{r\times K}$. Without loss of generality, we can assume that the first $r$ columns of $G'$ are linearly independent. Thus, we can obtain another optimal code $G$ from $G'$ which is systematic in the first $r$ columns, i.e., $G=[I|P_{r\times K-r}]$. Let $C$ be an involutory permutation matrix. Clearly the first $r$ columns of $G$ commute with $C$. Consider a  $L\times K$ $X$-fitting matrix $B_X^X$ obtained from $F_X$ by permuting the first $r$ columns according to $\sigma_C$, replacing all other ($K-r$) columns and the $1$s by $X$. 
By Theorem \ref{thm1}, the extended IC problem given by the fitting matrix $\begin{pmatrix} 
F_X & B_X^X \\
B_X^X & F_X 
\end{pmatrix}$ is a rank-invariant extension of $\cal I$ having the  index code $(G~CG)$.
\subsection{A class of special IC problems and their rank-invariant extensions}
\label{specialclass}
In Section \ref{subsecsimpleclass}, we obtain a simple rank-invariant extension of any index coding problem. By Theorem \ref{thm1}, the number of $r\times r$ submatrices of $G$ which commute with $C$ play a role in the structure of ${\cal I}_{ext}$. 
In the rest of this section, we construct another class of special IC problems satisfying conditions of Theorem \ref{thm1} and thus having rank-invariant $2$-order extensions. These IC problems have solutions consisting only of submatrices which commute with the chosen involutory permutation matrix $C$. We will invoke Lemma \ref{anothercommutingmatrix} for this purpose. 


Consider an IC problem ${\cal I}_{ABC}$ consisting of $rT$ messages (for some positive integers $r,T$), such that the messages can be divided into $T$ blocks of $r$ messages, denoted by ${\cal I}_i\subseteq [1:rT], i=1,...,T.$ The $r$ messages in a particular block ${\cal I}_i$ are referred to as $\{k_i:1\leq k\leq r\}$. The $T$ blocks are segregated into three types, $Type_A,~Type_B,$ or $Type_C$, having the following properties.

$\underline{Type_A~\text{blocks}}$: The following holds for all receivers $t$ which demand message $k_i$ in block ${\cal I}_i$, $\forall {\cal I}_i$ in $Type_A$, $\forall k_i\in {\cal I}_i$.  
\begin{itemize}
\item $k_j\in S(t)\cap {\cal I}_j$, $\forall {\cal I}_j\in Type_A, i\neq j$.
\item $\sigma_C(k)_j\in S(t)\cap {\cal I}_j$, $\forall {\cal I}_j\in Type_B$ (where $\sigma_C(k)_j$ is the message corresponding to the image of $k$ according to $\sigma_C$ in ${\cal I}_j$).
\item $k_j\in S(t)\cap {\cal I}_j$, $\forall {\cal I}_j\in Type_C.$
\item If $k$ is not a fixed point of $\sigma_C$ (i.e., there is a swap $(k,\sigma_C(k))$), then $\sigma_C(k)_j\in S(t)\cap{\cal I}_j$, $\forall {\cal I}_j\in Type_C.$
\end{itemize}

$\underline{Type_B~\text{blocks}}$: $Type_B$ blocks have the same properties as $Type_A$, except that `$Type_A$' in the above properties is swapped with `$Type_B$'.

$\underline{Type_C~\text{blocks}}$: $Type_C$ blocks satisfy the following properties, all receivers $t$ which demand message $k_i$ in block ${\cal I}_i$, $\forall {\cal I}_i$ in $Type_C$, $\forall k_i\in {\cal I}_i$. 
\begin{itemize}
\item $k_j\in {\cal I}_j\cap S(t),~\forall {\cal I}_j\in~Type_C$ such that $i\neq j$.
\item If $k$ is not a fixed point of $\sigma_C$ (i.e., there exists a swap $(k,\sigma_C(k))$), then $\sigma_C(k)_j\in {\cal I}_j\cap S(t), ~\forall {\cal I}_j\in~Type_C$ (including $j=i$), and either of the following conditions hold 
	\begin{enumerate}
	\item For \textit{any} ${\cal I}_j\in Type_A$, $k_j\in {\cal I}_j\cap S(t)$ \textit{and} for \textit{any} ${\cal I}_j\in Type_B$, $\sigma_C(k)_j\in{\cal I}_j\cap S(t)$. (OR)
	\item For \textit{any} ${\cal I}_j\in Type_A$, $\sigma_C(k)_j\in {\cal I}_j\cap S(t)$ \textit{and} for \textit{any} ${\cal I}_j\in Type_B$, $k_j\in{\cal I}_j\cap S(t)$.
	\end{enumerate}
\item If $k$ is a fixed point of $\sigma_C$, then both of the following conditions hold.
\begin{enumerate}
\item $k_j\in{\cal I}_j\cap S(t), \forall I_j\in Type_A$. 
\item $\sigma_C(k)_j\in{\cal I}_j\cap S(t), \forall I_j\in Type_B$. 
\end{enumerate}
\end{itemize} 

The reason for these properties for ${\cal I}_{ABC}$ will be clear to the reader upon reading the following proposition.
\begin{proposition}
\label{thm2}
The IC problem ${\cal I}_{ABC}$ has a $r\times rT$ index code $G$ such that
\[
 G^{{\cal I}_i}= 
\begin{cases}
    I,& \text{if}~{\cal I}_i\in~Type_A\\
    C,& \text{if}~{\cal I}_i\in~Type_B\\
		I+C-C_1,& \text{if}~{\cal I}_i\in~Type_C,
\end{cases}
\]
where $C_1$ is as in Lemma \ref{anothercommutingmatrix}, and $G^{{\cal I}_i}$ is the $r\times r$ submatrix of $G$ corresponding to the column indices in ${\cal I}_i$. 
\end{proposition}
\begin{IEEEproof}
We show that decoding holds for any arbitrary message in each type. 

$\underline{Type_A}$: Consider a message $k_i\in {\cal I}_i\in Type_A$ such that $k$ is a fixed point of $\sigma_C$. From the $k^{th}$ row of $G$, the $k^{th}$ transmitted symbol is 
\begin{eqnarray}
\label{eqn301}
x_{k_i}+\sum_{\substack{j\neq i\\ {\cal I}_j\in Type_A}}x_{k_j}+\sum_{{\cal I}_j\in Type_B}x_{\sigma_C(k)_j}+\sum_{{\cal I}_j\in Type_C}x_{k_j}.
\end{eqnarray}
The above expression is obtained by observing that all the other columns than those corresponding to the above messages in the $k^{th}$ row of $G$ have $0$s. By definition, any receiver demanding a message $k_i\in {\cal I}_i\in Type_A$ (such that $k$ is a fixed point of $\sigma_C$) contains all of the messages in (\ref{eqn301}) except $x_{k_i}$, as side-information. Hence $x_{k_i}$ is decodable. 

If $k$ is not a fixed point of $\sigma_C$, then the $k^{th}$ transmitted symbol is
\begin{eqnarray}
\nonumber
x_{k_i}+\sum_{\substack{j\neq i\\ {\cal I}_j\in Type_A}}x_{k_j}+\sum_{{\cal I}_j\in Type_B}x_{\sigma_C(k)_j}~~~~~~\\
\label{eqn302}
~~~~~~~~~~~+\sum_{{\cal I}_j\in Type_C}x_{k_j}+\sum_{{\cal I}_j\in Type_C}x_{\sigma_C(k)_j}.
\end{eqnarray}
Once again, by definition, any receiver demanding message $k_i\in{\cal I}_i\in Type_A$ (for $k$ not being a fixed point of $\sigma_C$) has all the messages in $(\ref{eqn302})$ (except for $x_{k_i}$) as side-information. Hence $x_{k_i}$ is decodable. 

$\underline{Type_B}$: Checking for decodability of $Type_B$ messages is similar to the case of messages in $Type_A$, hence we leave this to the reader. 

$\underline{Type_C}$: Consider a message $k_i\in{\cal I}_i\in Type_C$ such that $k$ is a fixed point of $\sigma_C$. The $k^{th}$ message transmission (corresponding to the $k^{th}$ row of $G$) can be written as 
\begin{eqnarray}
\label{eqn303}
x_{k_i}+\sum_{\substack{j\neq i\\ {\cal I}_j\in Type_C}}x_{k_j}+\sum_{{\cal I}_j\in Type_A}x_{k_j}+\sum_{{\cal I}_j\in Type_B}x_{\sigma_C(k)_j}.
\end{eqnarray}
An arbitrary receiver desiring $x_{k_i}$ can thus decode $x_{k_i}$ by using the side-information available (according to the definition of $Type_C$ messages). 

Suppose $k$ is not a fixed point. The $k^{th}$ transmission of the index code is 
\begin{eqnarray}
\nonumber
x_{k_i}+\sum_{\substack{j\neq i\\ {\cal I}_j\in Type_C}}x_{k_j}+\sum_{{\cal I}_j\in Type_C}x_{\sigma_C(k)_j}~~~~~~\\
\label{eqn304}
+\sum_{{\cal I}_j\in Type_A}x_{k_j}+\sum_{{\cal I}_j\in Type_B}x_{\sigma_C(k)_j}.
\end{eqnarray}
Thus, using (\ref{eqn304}), a receiver demanding $x_{k_i}\in{\cal I}_i$ at which condition 1) is satisfied can decode $x_{k_i}$. On the other hand, if the receiver satisfies condition 2), then it can use the $\sigma_C(k)^{th}$ transmission, which can be expressed as 
\begin{eqnarray}
\nonumber
x_{k_i}+\sum_{\substack{j\neq i\\ {\cal I}_j\in Type_C}}x_{k_j}+\sum_{{\cal I}_j\in Type_C}x_{\sigma_C(k)_j}~~~~~~\\
\label{eqn305}
+\sum_{{\cal I}_j\in Type_A}x_{\sigma_C(k)_j}+\sum_{{\cal I}_j\in Type_B}x_{k_j}.
\end{eqnarray}
\end{IEEEproof}
Using the IC problem ${\cal I}_{ABC}$ defined in Section \ref{specialclass} as the initial IC problem ${\cal I}$ in Theorem \ref{thm1}, we have the following corollary.
\begin{corollary}[Corollary to Theorem \ref{thm1}]
\label{corr1}
Let ${\cal I}={\cal I}_{ABC}$ with fitting matrix $F_X$ and $G$ as in Proposition \ref{thm2} be its index code. Let $B_X^X$ be obtained from $F_X$ according to Theorem \ref{thm1} (i.e., by permuting the columns in each ${\cal I}_i$ ($i\in[1:T]$) according to $\sigma_C$). Then  $(G~CG)$ is an index code for the extended IC problem ${\cal I}_{ext}$ with fitting matrix 
$\begin{pmatrix} 
F_X & B_X^X \\
B_X^X & F_X 
\end{pmatrix}.
$ Furthermore, if $G$ is optimal for ${\cal I}$, then $(G~CG)$ is an optimal code for ${\cal I}_{ext}$ and ${\cal I}_{ext}$ is a rank-invariant $2$-order extension of ${\cal I}$. 
\end{corollary}
\begin{example}
\label{rankunchanged}
Let $\sigma_C=(13)(2)$ and thus $C=\left(\begin{array}{ccc}
    0&0&1\\
    0&1&0\\
    1&0&0
\end{array}\right)$ be the involutory permutation matrix. Consider an IC problem ${\cal I}$ whose fitting matrix $F_X$ is given by
\[
\left(
\begin{array}{ccc|ccc|ccc|ccc}
 1 & 0 & 0 & 0 & 0 & X & 0 & 0 & X & X & 0 & X \\
 0 & 1 & 0 & 0 & X & 0 & 0 & X & 0 & \boxed{X} & X & 0 \\
 0 & 0 & 1 & X & 0 & 0 & X & 0 & 0 & X & 0 & X \\
 0 & \boxed{X} & X & 1 & 0 & 0 & X & 0 & 0 & X & 0 & X \\
 0 & X & 0 & 0 & 1 & 0 & 0 & X & 0 & 0 & X & \boxed{X} \\
 X & 0 & 0 & 0 & 0 & 1 & \boxed{X} & 0 & X & X & 0 & X \\
 0 & 0 & X & X & 0 & 0 & 1 & 0 & 0 & X & 0 & X \\
 0 & X & 0 & 0 & X & 0 & 0 & 1 & 0 & \boxed{X} & X & 0 \\
 X & 0 & 0 & 0 & 0 & X & 0 & 0 & 1 & X & 0 & X \\
 0 & 0 & X & X & 0 & 0 & X & 0 & 0 & 1 & 0 & X \\
 0 & X & 0 & 0 & X & \boxed{X} & 0 & X & 0 & 0 & 1 & 0 \\
 \undermat{Type_A}{X & 0 & 0}&\undermat{Type_B}{0 & 0 & X}&\undermat{Type_B}{0 & 0 & X}&\undermat{Type_C}{X & 0 & 1}\\
\end{array}
\right).
\]

\vspace{0.5cm}
Note that the side-information messages marked as $\boxed{X}$ indicate additional side-information at the receivers apart from those required in Section \ref{specialclass}. By Theorem \ref{thm2}, we have an index code for ${\cal I}$ as 
\[
G = \left(\underbrace{
\begin{array}{ccc}
    1&0&0\\
    0&1&0\\
    0&0&1
\end{array}}_{I}
\underbrace{
\begin{array}{ccc}
    0&0&1\\
    0&1&0\\
    1&0&0
\end{array}}_{C}
\underbrace{
\begin{array}{ccc}
    0&0&1\\
    0&1&0\\
    1&0&0
\end{array}}_{C}
\underbrace{
\begin{array}{ccc}
    1&0&1\\
    0&1&0\\
    1&0&1
\end{array}}_{I+C-C_1}
\right)
\]
where 
\[
C_1 = \left(
\begin{array}{ccc}
    0&0&0\\
    0&1&0\\
    0&0&0
\end{array}
\right).
\]
This code is also optimal, as it is easy to see from $F_X$ that $minrk(F_X)\geq 3$. Let $B_X^X$ be the matrix obtained applying Theorem \ref{thm2} to $F_X$ with ${\cal I}_1=\{1,2,3\}, {\cal I}_2=\{4,5,6\}, {\cal I}_3=\{7,8,9\}$ and ${\cal I}_4=\{10,11,12\}$, i.e., $B_X^X$ is given as 
\[
\left(
\begin{array}{ccc|ccc|ccc|ccc}
 0 & 0 & X & X & 0 & 0 & X & 0 & 0 & X & 0 & X \\
 0 & X & 0 & 0 & X & 0 & 0 & X & 0 & 0 & X & \boxed{X} \\
 X & 0 & 0 & 0 & 0 & X & 0 & 0 & X & X & 0 & X \\
 X & \boxed{X} & 0 & 0 & 0 & X & 0 & 0 & X & X & 0 & X \\
 0 & X & 0 & 0 & X & 0 & 0 & X & 0 & \boxed{X} & X & 0 \\
 0 & 0 & X & X & 0 & 0 & X & 0 & \boxed{X} & X & 0 & X \\
 X & 0 & 0 & 0 & 0 & X & 0 & 0 & X & X & 0 & X \\
 0 & X & 0 & 0 & X & 0 & 0 & X & 0 & 0 & X & \boxed{X} \\
 0 & 0 & X & X & 0 & 0 & X & 0 & 0 & X & 0 & X \\
 X & 0 & 0 & 0 & 0 & X & 0 & 0 & X & X & 0 & X \\
 0 & X & 0 & \boxed{X} & X & 0 & 0 & X & 0 & 0 & X & 0 \\
 0 & 0 & X & X & 0 & 0 & X & 0 & 0 & X & 0 & X \\
\end{array}
\right)
\]
By Corollary \ref{corr1}, $(G~CG)$ is a solution to the extended IC problem with fitting matrix $\begin{pmatrix} 
F_X & B_X^X \\
B_X^X & F_X 
\end{pmatrix}$.
\end{example} 

\end{document}